\documentclass[graybox,vecphys,natbib]{svmult}

\usepackage{amsmath,amssymb,bm,color}

\usepackage{mathptmx}       
\usepackage{helvet}         
\usepackage{type1cm}        
\usepackage{makeidx}         
\usepackage{graphicx}        
\usepackage{multicol}        
\usepackage{courier}        
\usepackage[bottom]{footmisc}

\makeindex             

\def\apj{ApJ}
\def\apjl{ApJ Lett.}

\def\aap{A\&A}
\def\aap{A\&A}
\def\mnras{Mon. Not. R. Astron. Soc.}
\def\nat{Nature}

\newcommand{\mbf}[1]{\ensuremath\bm{#1}}

\begin{document}

\title*{Combustion in thermonuclear supernova explosions}
\author{Friedrich K.~R{\"o}pke}
\institute{Heidelberger Institut f{\"u}r Theoretische Studien,
  Schloss-Wolfsbrunnenweg 35, 69118 Heidelberg,
  Germany\\ and\\ Zentrum f{\"u}r Astronomie der Universit{\"a}t
  Heidelberg, Institut f{\"u}r Theoretische Astrophysik,
  Philosophenweg 12, 69120 Heidelberg,
  Germany\\ \email{friedrich.roepke@h-its.org}}
%
%
\maketitle

\abstract{
Type Ia supernovae are associated with thermonuclear explosions of
white dwarf stars. Combustion processes convert material in nuclear
reactions and release the energy required to explode the stars. At the
same time, they produce the radioactive species that power radiation
and give rise to the formation of the observables. Therefore, the
physical mechanism of the combustion processes, as reviewed here, is
the key to understand these astrophysical events.\\
Theory establishes two distinct modes of propagation for combustion
fronts: subsonic deflagrations and supersonic detonations. Both are
assumed to play an important role in thermonuclear supernovae.  The
physical nature and theoretical models of deflagrations and
detonations are discussed together with numerical implementations.  A
particular challenge arises due to the wide range of spatial scales
involved in these phenomena. Neither the combustion waves nor their
interaction with fluid flow and instabilities can be directly resolved
in simulations. Substantial modeling effort is required to
consistently capture such effects and the corresponding techniques are
discussed in detail. They form the basis of modern multidimensional
hydrodynamical simulations of thermonuclear supernova explosions. The
problem of deflagration-to-detonation transitions in thermonuclear
supernova explosions is briefly mentioned.
}

\section{Introduction}  
\label{intro}   

Theory associates Type Ia supernovae\index{supernovae, Type Ia} with
thermonuclear explosions of white dwarf stars. Therefore, any model of
these spectacular astrophysical events is based on describing the
physical processes leading to such explosions. Thermonuclear reactions
release the energy necessary to overcome the gravitational binding of
the star. What are these reactions? How do they spread over the
stellar material after ignition? These questions are answered by
combustion theory which was extensively developed for chemical
terrestrial processes because of their importance for technical
applications. Over the past two decades, there has been an increased
interest in modeling the combustion processes in thermonuclear
supernovae. As we will discuss below, the phenomena known from
chemical combustion find many counterparts here. This led to a rapid
improvement of multidimensional simulations of thermonuclear supernova
explosions that greatly benefited from knowledge and techniques in
technical combustion modeling. Some of the regimes of turbulent
combustion that we will identify in the context of supernova modeling
are encountered, for instance, in car engines, and are well known and
studied there.

We will argue in the following sections that thermonuclear
burning\index{thermonuclear burning} in Type Ia supernovae propagates
as combustion waves subject to various instabilities. This makes the
explosion physics of thermonuclear supernovae rich and its numerical
simulation challenging. A variety of physical effects determine the
propagation of such combustion waves, and the scales of interaction
range from microscopic lengths below millimeters up to the radius of
the exploding stars (thousands of kilometers). Modeling thermonuclear
combustion in white dwarf stars is therefore a demanding multi-scale
multi-physics problem.

\section{Combustion in White Dwarf matter}

The phenomenon of combustion\index{combustion in white dwarf matter}
is generally modeled as an interplay of fluid mechanics and reactions:
reactions convert species and release the energy that drives the
process. This causes fluid flow and energy transport that propagate
the burning into fuel material. All this is captured in a set of
partial differential equations that result from combining the
Navier-Stokes equations\index{Navier-Stokes equations} of fluid
dynamics\index{fluid dynamics} with reactive and other source terms
and additional terms that account for energy transport by diffusion
and conduction. The equations are based on the concepts of mass
conservation
  \begin{equation}
    \label{grf_mass}
    \frac{\partial \rho}{\partial t} + \mbf{\nabla} \cdot (\rho
    \mbf{v}) = 0,
  \end{equation}
   momentum balance
  \begin{equation}
    \label{grf_momentum}
    \frac{\partial \rho \mbf{v}}{\partial t} + \mbf{\nabla} \cdot (\rho
    \mbf{vv}) + \mbf{\nabla} \cdot \mbf{\Pi} = \rho \mbf{f},
  \end{equation}
   species balance\footnote{Note that the combination of mass conservation
     with the equations describing balance of species overdetermines
     the system and thus Eqs.~(\ref{grf_mass}) and (\ref{grf_species})
     have to be treated in a consistent way.}
  \begin{equation}
    \label{grf_species}
    \frac{\partial \rho X_i}{\partial t} + \mbf{\nabla} \cdot
    (\rho X_i \mbf{v})  = - \mbf{\nabla} \cdot (\rho
    \mbf{v}^{\mathrm{D}}_i X_i) + \rho \omega_{X_i}
     \qquad i=1\ldots N,
  \end{equation}
   and energy balance
  \begin{equation}
    \label{grf_energy}
    \frac{\partial \rho e_\mathrm{tot}}{\partial t} + \mbf{\nabla} \cdot
    (\rho e_\mathrm{tot} \mbf{v}) + \mbf{\nabla} \cdot ( \mbf{v}
    \mbf{\Pi}) = \rho \mbf{v} \cdot \mbf{f} + \rho \sum
    \limits_{i=1}^{N} X_i \mbf{v}_i^\mathrm{D} \cdot \mbf{f}_i
    - \mbf{\nabla} \cdot \mbf{q} + \rho S.
  \end{equation}
Here, $\rho$, $\mbf{v}$, $e_\mathrm{tot}$, $\mbf{\Pi}$, $\mbf{f}$ ,
$X_i$ , $\mbf{v}^\mathrm{D}$ , $\omega$, $\mbf{q}$, and $S$ denote mass
density, fluid velocity, specific (i.e.\ per unit mass) total energy,
pressure tensor, external forces, mass fraction of species $i$ (index
$i$ running from $1$ to the number of considered species $N$),
diffusion velocity, reaction rate, heat flux, and energy source terms
due to reactions, respectively.

Eq.~(\ref{grf_mass}) equals the temporal change of the mass density to
the divergence of the mass flux density (the sign is
convention). Integrated over a control volume, Eq.~(\ref{grf_mass})
states that mass can only change inside it by a mass flux over its
surface. Species, momentum, and energy are not strictly conserved. In
addition to changes due to fluxes over the surface, they can be
produced or destroyed by physical mechanisms inside the volume. This
is captured in the source terms on the right-hand sides of
Eqs.~(\ref{grf_species}) -- (\ref{grf_energy}): momentum can be
changed under the action of an external force, species are changed by
diffusion and reactions, and energy is affected by external forces,
diffusion processes, heat flux, and energy release or consumption in
reactions.  The list of source terms included in
Eqs.~(\ref{grf_momentum}) -- (\ref{grf_energy}) is not exhaustive. As
pointed out by \cite{timmes1992a}, radiative heat flux is unimportant
in white dwarf matter, and we therefore neglect it here. The potential
dynamic effects of magnetic fields are also not considered. Our choice
of effects accounted for in the above equations, however, allows for a
model that captures the basics of combustion physics in thermonuclear
supernovae.

The system of Equations (\ref{grf_mass}) -- (\ref{grf_energy}) is
capable of describing complex physical processes. In particular, we
emphasize the importance of the second term on the right-hand side of
Eq.~(\ref{grf_momentum}), the divergence of the momentum flux density,
often referred to as ``advection term.'' This term is nonlinear in
the fluid velocity leading to the rich (and theoretically still not
satisfactorily understood) phenomenon of turbulence. Its impact on
combustion processes will be discussed in Sect.~\ref{sect:def_inst}.
The ratio of the magnitude of the advection term to that of viscosity
is measured by the Reynolds number $\mathit{Re}$\index{Reynolds
  number} on a certain spatial scale. In the situations encountered
for thermonuclear burning in white dwarf stars, the value of the
Reynolds number is huge (it can reach $\mathit{Re} \sim 10^{14}$ on
length scales comparable to that of the white dwarf star). To a good
approximation, viscosity effects can be neglected in models that
consider such astrophysical combustion processes on large scales. The
pressure tensor then reduces to the hydrodynamical pressure $P$, and
Eq.~(\ref{grf_momentum}) simplifies to the Euler equation.

To close the system of Eqs.~(\ref{grf_mass}) -- (\ref{grf_energy}), an
appropriate equation of state\index{equation of state, white dwarf
  matter} has to be provided. It relates pressure to other quantities
such as mass density, internal energy, and composition. White dwarf
stars are compact objects in which matter is fully ionized.  For white
dwarf matter, the ions dominate the mass density and are treated as an
ideal gas. The electrons, however, provide the main contributions to
pressure and energy. They form a degenerate gas and may be
relativistic to a variable degree. Additional effects due to
radiation, electron-positron pairs, and Coulomb interactions may also
be incorporated in the equation of state. Unlike the ideal gas
equation of state widely used in computational fluid dynamics,
astrophysical equations of state are often provided in form of tables
rather than as a closed expression. In addition, relations modeling
heat flux and diffusion have to be specified. In the case of
thermonuclear supernova explosions, the most important external force
is gravity. The corresponding terms in the above equations can be
determined from solving the Poisson equation for the given mass
distribution.

Reaction rates and the energy source term due to reactions depend on
temperature, density, and composition,
\begin{eqnarray*}
  \omega_{X_i} &=& \omega_{X_i}(\rho, T, X_i),\\
  S &=& S(\omega_{X_i}).
\end{eqnarray*}
The reactions in a combustion system can be chemical in nature, as is
usually the case in terrestrial technical applications, or nuclear, as
in Type Ia supernovae. Here, we are interested in the latter
case. Nuclear reactions are formally treated with a nuclear reaction
network \citep[e.g.][]{hix2006a}\index{nuclear reaction network}. This
set of coupled ordinary differential equations describes the temporal
change of the abundance of each species $i$ due to electron captures,
$\beta$-decays, photodisintegrations, two-, and three-body
reactions (for higher-order reactions, the probability is low and they
can be neglected). These effects depend on temperature, density, and
the abundances of the other species in the system.

The numerical implementation of the hydrodynamics solver in
thermonuclear supernova explosion simulations usually follows finite
volume techniques. For a textbook on that matter we refer to
\citet{toro2009a}.

Combustion is encountered in many situations in nature and
technology. The characteristics of such processes, however, vary
widely. Chemical combustion is classified depending on how the
reactants are brought together in the fuel. In \emph{premixed
  combustion}, all necessary ingredients for the reaction are contained
in the fuel mixture, whereas diffusion processes bring the reactants
into contact in \emph{non-premixed} combustion. Both types are
discussed in detail in \cite{peters2000a}. In thermonuclear
supernovae, the fuel material contains all species required to set off
nuclear burning, and therefore this situation resembles that of
premixed combustion, which we will focus on in the following
discussion.

\section{Combustion fronts}
\label{sect:fronts}

Reactions in combustion systems are usually initiated by an ignition
in a certain small volume from which they spread out. In the case of
thermonuclear supernovae, the extremely strong dependence of the
associated thermonuclear reaction rates on temperature localizes the
burning in space so that it propagates as a combustion
wave\index{combustion waves}. For example, the reaction rate of carbon
fusion, which initiates thermonuclear burning in carbon-oxygen white
dwarf stars, scales at $T \sim 10^{10}\, \mathrm{K}$ as $10^{12}$
\citep{hansen1994a}. Therefore, burning peaks sharply at the places of
highest temperature, and the reaction is confined to a thin layer
in space that propagates through the material. At high fuel densities,
the width of such combustion waves may be tiny fractions of a
millimeter. At lower densities it broadens, but it still remains small
compared to the scales of the exploding white dwarf star with a radius
of thousands of kilometers. The idealized treatment of a combustion
wave as a sharp front, where fuel is instantaneously converted to ash
and the corresponding energy is released, thus provides an excellent
approximation for combustion modeling on these large scales.

This \emph{discontinuity approximation}\index{combustion fronts,
  discontinuity approximation} establishes the simplest model of a
combustion wave. Its internal structure is not resolved, but the
correct relation between the hydrodynamical states ahead and behind
the combustion front is recovered. The discontinuity approximation
simplifies the equations describing the system: microphysical
transport due to diffusion and heat conduction can be neglected. If
the fluid can be treated as ideal and in the absence of gravity and
other external forces, the system reduces to the Euler equations with
source terms due to reactions.

Discontinuities are not among the solutions of the system as stated
above. This differential form of the equations of fluid dynamics
allows only for continuous (``strong'') solutions. The integral form,
however, also allows for discontinuous (``weak'') solutions. If
treated in the differential from, discontinuous solutions require the
introduction of additional jump conditions\index{jump conditions}
relating the hydrodynamical states on both sides of the
discontinuity. Neglecting reactions, conservation of mass, momentum,
and energy over discontinuities require continuous fluxes of these
quantities over the discontinuity (which may carry jumps in other
quantities such as density, temperature, etc.).

For simplicity, consider the system in one spatial dimension in the
rest frame of the combustion front, and let $u$ denote the velocity
component normal to it.  Combining the continuity of mass flux
density, $M := \rho u$, and momentum flux density, $\rho u^2 + P$,
over the front [the corresponding expressions can easily be read off
  from the terms on the left-hand side of Eqs.~(\ref{grf_mass}) and
  (\ref{grf_momentum})], we arrive at the \emph{Rayleigh criterion}
\citep[see][]{landau1987a} relating the unburnt and burnt states
(denoted with subscripts ``u'' and ``b'', respectively):
\begin{equation}
\label{eq:rayleigh}
- M^2 = - \left( \rho_\mathrm{u} u_\mathrm{u}\right)^2 = - \left(
\rho_\mathrm{b} u_\mathrm{b}\right)^2 = \frac{P_\mathrm{b} -
  P_\mathrm{u}}{V_\mathrm{b} - V_\mathrm{u}}.
\end{equation}
The specific volume is defined as $V :=
1/\rho$. Eq.~(\ref{eq:rayleigh}) describes a straight line in the
$P$-$V$-plane\index{Rayleigh line}. It must have a negative slope
corresponding to $-M^2$, because the mass flux cannot be
imaginary. The burnt state is characterized by a density and a
pressure that is connected to the unburnt state by this line.

The Rayleigh criterion (\ref{eq:rayleigh}) derives from the principles
of mass and momentum conservation and confines the \emph{mechanically
  possible} burnt state to a line in the $P$-$V$-plane. A second
relation fixes the possible burnt states to a point. If no energy were
released in the combustion wave, the energy flux $(\rho e_\mathrm{tot}
+ P)u = (1/2 u^2 + e_\mathrm{int} + P/\rho) \rho u $ over it would be
continuous. Here, we introduce the specific internal energy
$e_\mathrm{int} := e_\mathrm{tot} - u^2/2$. This results in a
thermodynamic condition for the ``burnt'' state, the so-called
\emph{Hugoniot adiabat}\index{Hugoniot adiabat}
\citep[see][]{landau1987a}:
\begin{equation}
 e_\mathrm{int, u} - e_\mathrm{int, b} = - \frac{P_\mathrm{u} -
   P_\mathrm{b}}{2} \left( V_\mathrm{b} - V_\mathrm{u}\right).
\end{equation}
Of course, no energy release in the burning is accounted for up to
now, so it would be more correct to speak of the
``post-discontinuity'' state here; this is not done to avoid excessive
introduction of notation.  For common equations of state, the Hugoniot
adiabat is a parabola in the $P$-$V$-plane connecting the unburnt
state to all \emph{thermodynamically allowed} ``burnt'' states. It is
important to note that the slope of the tangent to the Hugoniot
adiabat at a certain point measures the speed of sound in the
corresponding state, while the slope of the Rayleigh line through that
same point represents the velocity of the front with respect to that
state \citep[see][]{landau1987a}. The curve's intersection with the
Rayleigh line marks the \emph{physically realized} (i.e. both
mechanically and thermodynamically admissible) ``burnt'' state
$(V_\mathrm{b}, P_\mathrm{b})$.

\begin{figure}[t]
\centerline{\includegraphics[width=0.6\textwidth]{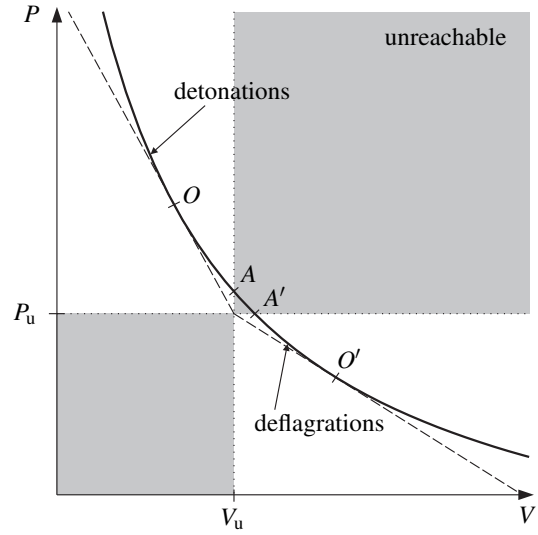}}
\caption{Relation between unburnt and burnt states in the
  discontinuity approximation of combustion. The detonation adiabat is
  plotted as a solid curve, and the tangential Rayleigh lines are
  dashed. [See also \cite{landau1987a} and \cite{linan1993a}. Reprinted
  with minor modifications from \cite{roepke_phd}]\label{fig:hugoniot}
}
\end{figure}

The process modeled so far without energy release over the combustion
front corresponds to a hydrodynamical shock. If energy release is
considered, the Hugoniot curve shifts upward in the $P$-$V$-plane and
does not contain the unburnt state any longer:
\begin{equation}
 e_\mathrm{int, u} - e_\mathrm{int, b} = \Delta h_0 -
 \frac{P_\mathrm{u} - P_\mathrm{b}}{2} \left( V_\mathrm{b} -
 V_\mathrm{u}\right).
\end{equation}
Here, $\Delta h_0$ denotes the difference in the formation enthalpies
of the burnt and unburnt material. This situation is illustrated in
Fig.~\ref{fig:hugoniot}. The shifted Hugoniot curve is usually
referred to as \emph{detonation adiabat}\index{detonation adiabat}. It
can intersect with Rayleigh lines issuing from the unburnt state
$(V_\mathrm{u}, P_\mathrm{u})$ for different values of $M$. Since
these lines must have a negative slope, the possible burnt state is
located on one branch of the detonation adiabat, either above point
$A$ or below point $A'$ in Fig.~\ref{fig:hugoniot}. Combustion in
these two branches is caused by different physical mechanisms as we
will discuss below.

The speed of sound in the unburnt material is graphical represented by
slope of the tangent to the Hugoniot adiabat in point $(V_\mathrm{u},
P_\mathrm{u})$. By construction (see Fig.~\ref{fig:hugoniot}), all
Rayleigh lines connecting $(V_\mathrm{u}, P_\mathrm{u})$ to the upper
branch of the upward-shifted detonation adiabat have a steeper slope,
which, in turn, measures the velocity of the combustion wave with
respect to the unburnt state \cite{landau1987a}. This means that all
processes leading to burnt states on the upper branch of the
detonation adiabat are caused by combustion fronts that propagate
supersonically with respect to the fuel. These are called
\emph{detonations}\index{detonations}. The converse holds for burnt
states on the lower branch of the detonation adiabat, and the
corresponding subsonically propagating combustion fronts are called
\emph{deflagrations}\index{deflagrations}.

The Rayleigh lines from $(V_\mathrm{u}, P_\mathrm{u})$ to the points
$O$ and $O'$ in Fig.~\ref{fig:hugoniot} mark a special case and are
indicated with dashed lines in the figure. They are tangents to the
detonation adiabat. With the same lines of arguments as above, one can
show \cite{landau1987a} that combustion waves terminating in points
$O$ and $O'$ propagate with the speed of sound in the burnt
material. These are the so-called \emph{Chapman-Jouguet}
points\index{Chapman-Jouguet points}. Above these points, the velocity
of the front is subsonic with respect to the burnt state; below it is
supersonic.

\section{Deflagrations in White Dwarf matter}

The simplified picture of a combustion front modeled as a
discontinuity discussed in Sect.~\ref{sect:fronts} provides useful
relations between burnt and unburnt states that directly derive from
the balance laws of fluid dynamics. It does not, however, address the
question of the physical processes that lead to the phenomenon of
propagating combustion waves on a microscopic level.

In contrast to detonations (see Sect.~\ref{sect:det}), deflagrations
propagate due to microphysical transport processes\index{deflagrations
  in white dwarf matter}. Energy is released by reactions in the
burning zone. Conduction and diffusion lead to a heating of the fuel
ahead of this zone so that it also reaches conditions for burning. In
this sense, deflagrations resemble the picture of a propagating
\emph{flame front}. Its velocity is given by the microphysical
transport processes and therefore slow compared to detonations that
will be discussed below.

The full structure of deflagration waves results from solving
Eqs.~(\ref{grf_mass}) -- (\ref{grf_energy}). The reaction zone,
where the reaction rate is significant, typically occupies only a
small fraction of the width of the structure, and it trails a more
extended preheat zone.

Deflagrations in white dwarf matter are more complex than this
schematic picture suggests. To solve the equation of combustion
hydrodynamics, the relevant microphysical processes have to be
modeled. Nuclear burning proceeds in a multitude of reactions,
starting out with carbon burning, over oxygen burning to silicon
burning, finally reaching nuclear statistical equilibrium, provided
the fuel densities are sufficiently high. The relevant reactions have
to be identified, and the corresponding reaction rates have to be
provided. In addition, energy transport has to be accounted for. It
turns out that the most important of these effects is thermal
conduction, an energy flux caused by a temperature gradient:
\begin{equation}
\mbf{q} = - \sigma \mbf{\nabla} {T}.
\end{equation}
In white dwarf matter, different energy carriers are present,
e.g., electrons, ions, and photons. The transport properties of these
carriers have to be accounted for in determining the thermal
conductivity $\sigma$. \cite{timmes1992a} discuss the corresponding
contributions and conclude that electron conduction dominates over all
other effects under conditions typical for white dwarf stars. The
reason is that final states for electronic scattering processes are
occupied below the difference of Fermi energy $E_\mathrm{F}$ of the
electron gas, which is $\sim$$1\, \mathrm{MeV}$, and the thermal
energy $k_\mathrm{b}T \sim 10 \, \mathrm{keV}$. This implies extremely
large mean free paths of the electrons leading to high values of
$\sigma$.

\begin{figure}[t]
\begin{center}
\includegraphics[width=0.8\textwidth]{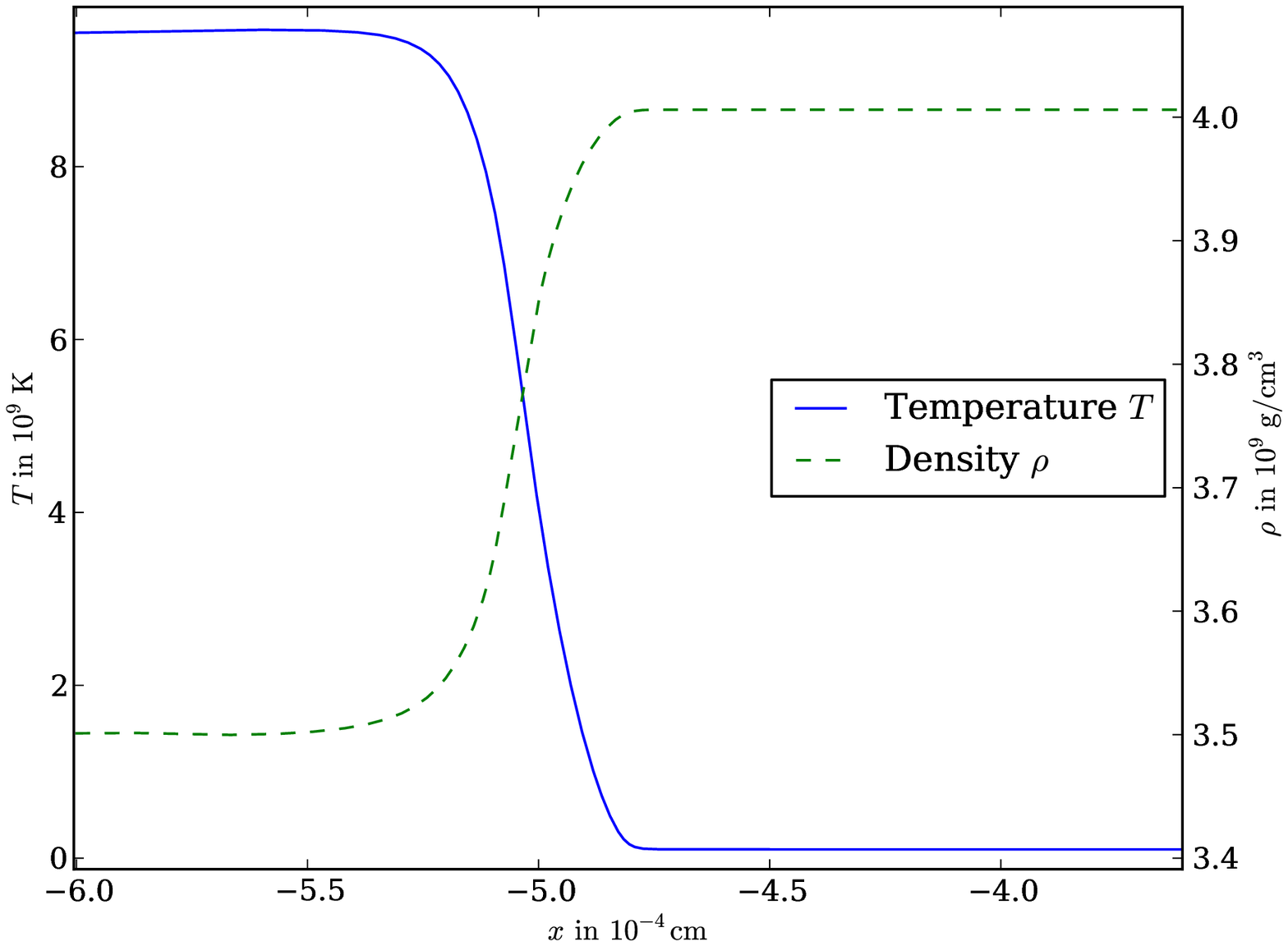}\\
\includegraphics[width=0.8\textwidth]{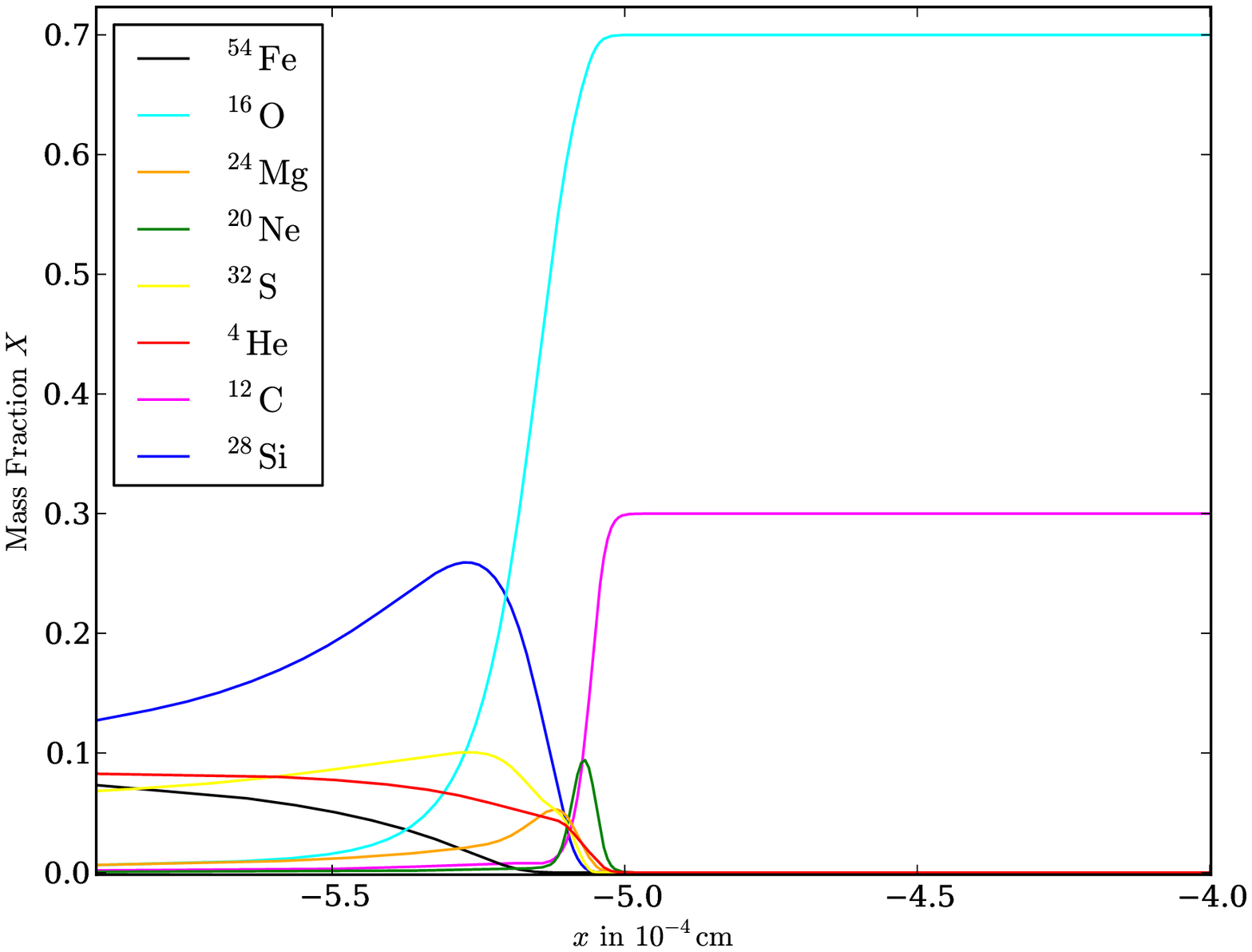}
\end{center}
\caption{ Structure of a deflagration flame for fuel at $4 \times 10^9
  \, \mathrm{g}\,\mathrm{cm}^{-3}$ with a composition of 30\% carbon
  and 70\% oxygen. Here, the flame advances from left to
  right. [Reprinted from the diploma thesis of Philipp Edelmann,
  Technische Universit{\"a}t M{\"u}nchen (2010), with permission of
  the author]
\label{fig:deflagration}}
\end{figure}

An important consequence is observed when comparing the magnitudes of
thermal conduction and diffusive energy transport, a ratio that
defines the Lewis number\index{Lewis number}. Typical values for white
dwarf matter are $\sim$$10^7$ \citep{timmes1992a}, while the Lewis
number is order of unity in terrestrial combustion processes. The
Prandtl number\index{Prandtl number} compares viscous transport to
thermal conduction, and its value is typically very small in white
dwarf matter \citep{timmes1992a}. These figures of merit reflect the
efficiency of electron conduction due to the high degeneracy of white
dwarf matter. This situation greatly simplifies the solution of the
equations of combustion hydrodynamics (\ref{grf_mass}) --
(\ref{grf_energy}). The huge Lewis number indicates that diffusive
processes are unimportant for the energy transport and can be
neglected. The low value of the Prandtl number demonstrates the
subdominance of viscous effects. Therefore, a model neglecting
diffusive transport and based on the Euler equations of hydrodynamics
rather than on the Navier-Stokes equations is justified. Another
argument for this approach is that the full set of equations can only
be solved numerically. Hydrodynamics solvers, however, introduce a
considerable numerical viscosity, and explicitly accounting for small
physical viscosities is pointless.

Numerical solutions of the resulting system determine the laminar
(i.e. in case of a planar front in the absence of any geometrical
perturbation) propagation speed $s_\mathrm{l}$ of deflagration waves
as a function of fuel density, temperature, and composition. Results
are given by \cite{timmes1992a}. The different reactions occurring in
a deflagration in white dwarf matter are illustrated in
Fig.~\ref{fig:deflagration}. It shows the density, the temperature, and
the abundances of different species inside the structure. Note that the
thermodynamical structure evolves monotonically, although the various
reactions give rise to burning on different length scales due to
differences in their reaction rate. Therefore, the width of the
``carbon flame'' is almost an order of magnitude smaller than that of
the ``oxygen flame.''

\section{Deflagration Instabilities and interaction with turbulence}
\label{sect:def_inst}

In more than one spatial dimension, subsonic deflagrations are subject
to various instabilities. Some of them are relevant or even
fundamental for flame propagation in thermonuclear
supernovae. Deflagrations in Chandrasekhar-mass white dwarfs are one
prominent example. They can potentially unbind the stars
\citep[e.g.][]{reinecke2002d, gamezo2003a, roepke2007c}, or at least
lead to the ejection of a substantial part of their masses
\citep[][]{jordan2012b, kromer2013a, fink2014a}. This is not trivial
as subsonic deflagrations significantly expand of the fuel material
ahead of them due to their energy release. This drop in fuel density
eventually inhibits further burning, and there is a competition
between the star's expansion and the fuel consumption by the flame. A
simple laminar flame propagates too slowly. It does not burn
sufficient amounts of material to explode the star. The required
efficiency in energy release and species conversion is only possible
due to the action of instabilities\index{instabilities of deflagration
  waves} causing turbulence\index{turbulence} with which the flame
interacts and thus accelerates\index{deflagrations, turbulent}. We
will discuss below that instabilities are an inevitable physical
consequence of any multidimensional treatment of fluid dynamics and
flame propagation. These effects can only be parametrized in
one-dimensional models, and thus a multidimensional simulations are
mandatory when a solid and predictive explosion model is aimed for.

Not all potential flame instabilities are relevant for the case of
thermonuclear burning in supernovae. The diffusive-thermal instability
\citep{barenblatt1962a}, for instance, is suppressed at the prevailing
high Lewis numbers. The three main instabilities that potentially
affect the flame propagation are the Rayleigh-Taylor instability, the
Kelvin-Helmholtz instability, and the Landau-Darrieus instability. The
first two are general fluid flow instabilities, while the latter is
specific to burning fronts.

The Rayleigh-Taylor instability\index{Rayleigh-Taylor instability} is
caused by buoyancy in the corresponding supernova explosion
scenarios. The flame ignites near the stellar center and burns toward
the surface. Behind the flame, energy is released by nuclear
reactions. This partially lifts the degeneracy, and therefore the
density in the ashes is lower than in the fuel. The result is an
inverse density stratification in the gravitational field of the star,
which is buoyancy unstable. Perturbations at the interface between
fuel and ash grow. In the nonlinear stage, the Rayleigh-Taylor
instability leads to the formation of bubbles of hot ash material that
rise into the cold fuel. In between, downdrafts transport fuel
material toward the star's center. These counterflows at the fuel-ash
interface lead to shear at the flame. The Kelvin-Helmholtz
instability\index{Kelvin-Helmholtz instability} amplifies initial
perturbations and forms a wave pattern that grows into eddies in the
nonlinear stage.

The flame itself, however, is subject to an instability that arises
from its self-propagation. This so-called Landau-Darrieus
instability\index{Landau-Darrieus instability} is active on all scales
that are significantly larger than the internal flame width and
results from a refraction of stream lines in the vicinity of a flame
front due to the change in density over it. While the tangential
velocity component over the front is steady, mass flux conservation
requires a jump in the normal component. This widens streamlines in
the vicinity of bulges of the flame and consequently the fluid
velocity is lower locally. The laminar burning speed is thus larger
than the local fluid velocity and this causes bulges to
increase. Recesses become deeper by the opposite effect. Thus
perturbations deforming the flame front from planar geometry grow.

In the nonlinear regime, however, the flame stabilizes into a
cellular pattern \citep{zeldovich1966a}. Such a stabilization was
shown to be effective for thermonuclear flames in white dwarf matter
\citep{roepke2003a, roepke2004a, roepke2004b}. The Landau-Darrieus
instability is therefore not expected to have a significant impact on
burning in thermonuclear supernovae.

The Rayleigh-Taylor and Kelvin-Helmholtz instabilities, in contrast,
show no such nonlinear stabilization. They are fundamental for the
evolution of flame structure on the scales of the exploding white
dwarf star and lead to a considerable acceleration of burning in
thermonuclear supernovae. Both instabilities, however, do not act on
the smallest spatial scales. The growth time of the Rayleigh-Taylor
instability competes with the burning time scale. At the smallest
scales front distortions grow so slowly that they will be overrun by
the flame. As the growth time of the buoyancy-driven instability
increases with the length scale of the perturbations, there exists a
minimum scale for the Rayleigh-Taylor instability develop in the
presence of a self-propagating flame. This so-called fire-polishing
length can be estimated as \citep[see also][]{timmes1992a}
\begin{equation}
\lambda_\mathrm{fp} = \frac{g s_\mathrm{l}^2}{2 \pi}
\frac{\rho_\mathrm{u} - \rho_{b} }{\rho_{u} + \rho_\mathrm{b}},
\end{equation}
with $g$ denoting the gravitational acceleration.

The Kelvin-Helmholtz instability acts as a secondary effect, because
the shear motions necessary to trigger it are produced by the
large-scale uprising plumes of hot ashes in the nonlinear stage of
Rayleigh-Taylor instability. In its nonlinear regime, the
Kelvin-Helmholtz instability results in vortices that develop in the
shear region. A prerequisite for a Kelvin-Helmholtz unstable
configuration is a tangential discontinuity without a flow over
it. This is clearly not the case encountered for burning fronts. The
finite mass flow over them stabilizes flames against the
Kelvin-Helmholtz instability. Its effect is similar to that of
viscosity in shear layers leading to some stabilization. With buoyancy
acting to form fast-rising bubbles of ash, however, the situation
changes. The resulting shear velocities are much larger than the flow
velocities over the flame fronts. Indeed, in numerical simulations
\cite{niemeyer1997a} found that the flame becomes subject to the
Kelvin-Helmholtz instability once the shear velocity exceeds the
laminar burning speed. This is typically the case for thermonuclear
deflagrations in white dwarf stars.

The action of the Kelvin-Helmholtz instability is the primary cause
for the generation of turbulence. The importance of this effect can be
estimated from the Reynolds numbers of the shear flows around rising
bubbles of burning material that are as high as $10^{14}$. This gives
rise to the following picture: turbulent eddies are generated by shear
instability at the location of the flame at the large scales of the
buoyant plumes of burnt material. These eddies themselves are unstable
and decay to smaller eddies, thus establishing a turbulent cascade. In
this cascade, turbulent kinetic energy injected at the largest scales
is transported without loss through the so-called inertial range. The
velocity of the turbulent eddies steadily decreases toward smaller
scales. At some microscopic scale (much below a millimeter), the local
(scale-dependent) Reynolds number drops to $\sim$$1$ indicating that
viscous effects become important. At this so-called Kolmogorov scale
they dissipate the turbulent kinetic energy into heat.

\begin{figure}[t]
\centerline{\includegraphics[width=0.75\textwidth]{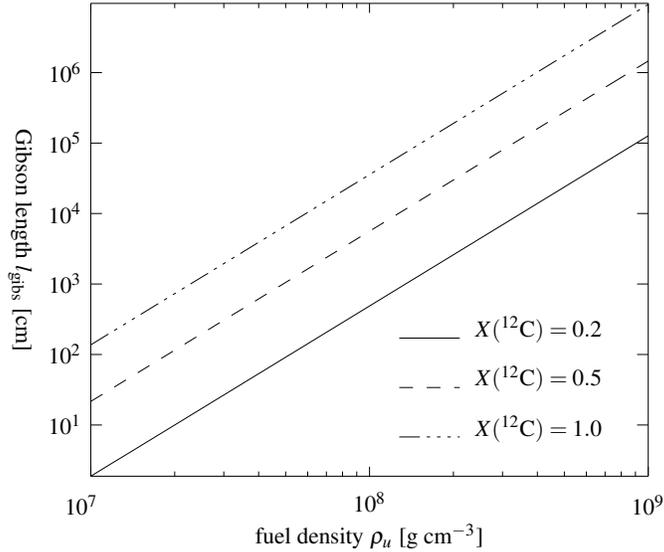}}
\caption{Gibson scale for typical deflagrations in Chandrasekhar-mass
  white dwarf stars as a function of fuel density for different carbon
  mass fractions in the white dwarf material. [Reprinted from from
  \cite{roepke_phd}]\label{fig:gibson} }
\end{figure}

The flame interacts with turbulent eddies on a wide range of
scales\index{turbulent combustion}\index{turbulent deflagrations}. On
scales much larger than its internal width, they deform and corrugate
the flame front. The scale down to which this effect is active depends
on the flame speed. Similar to the argument of the fire polishing
length for buoyancy instability, there should be a scale on which the
turbulent eddy velocities have become so low that they are comparable
to the laminar flame speed. This implies that below this so-called
\emph{Gibson scale}\index{Gibson scale} $l_\mathrm{Gibs}$ the flame
burns faster through the turbulent eddies than they turn over, and the
flame is nearly undistorted by the action of eddies. For constant
turbulent velocities, the size of the Gibson scale depends on the fuel
density. With lower fuel density the flame speed decreases and the
Gibson scale becomes smaller (see Fig.~\ref{fig:gibson}).

One important question is how far down in scale space the effect of
turbulent velocity fluctuations affects the flame, i.e. the size of
the Gibson scale. If this scale is significantly above the width of
the flame, turbulence will not interact with the internal flame
structure. The only effect is then a corrugation of the flame front on
large scales. This is the so-called \emph{flamelet
  regime}\index{turbulent combustion, flamelet regime} of turbulent
combustion \citep{peters2000a}.  While the microphysics of the burning
is unaffected and the laminar flame speed does not change, the overall
flame wrinkling has significant effect on the overall burning
efficiency. The flame surface is enlarged and therefore the net
burning rate increases.

If, however, the Gibson scale is smaller than the internal flame
width, turbulent eddies will affect the internal flame structure. They
can now transport material into and out of the flame structure and
thus disrupt it. This is the so-called \emph{distributed burning
  regime}\index{turbulent combustion, distributed burning
  regime}. Depending on the size of the Gibson scale, different
sub-regimes are distinguished \citep[see][for an extensive
  discussion]{peters2000a}: In the \emph{thin reaction zone regime},
turbulence will alter the structure of the preheat zone, but not the
reaction zone itself. When the Gibson scale reaches down to the width
of the actual reaction zone, burning is said to take place in the
\emph{broken reaction zones regime} and ultimately turbulence
completely dominates the burning process and spreads it out over space
in the \emph{well-stirred reactor regime}.

In a thermonuclear supernova explosion resulting from a deflagration,
the flame will enter the distributed burning regime. Towards the end
of the burning, turbulence freezes out and the gravitationally unbound
material approaches homologous expansion (\cite{roepke2005c}). Before
this happens, however, the flame will transition from the flamelet
regime of turbulent combustion to the distributed burning regime. As
the exploding white dwarf expands and the flame burns towards its
surface, the fuel density decreases thus reducing the Gibson scale
(see Fig.~\ref{fig:gibson}). At the same time, the internal width of
the deflagration grows with decreasing fuel density
\citep[e.g.][]{timmes1992a}. Once the flame reaches material with
$\lesssim$$10^7 \, \mathrm{g} \, \mathrm{cm}^{-3}$, the Gibson scale
falls below the flame width, and turbulence affects its internal
structure.

Overall, the picture is that deflagrations in thermonuclear supernova
explosions are far from propagating laminar. They possess a pronounced
multidimensional structure, being wrinkled by instabilities on large
scales, and the thus induced turbulence interacts with them on a wide
range of scales. A one-dimensional model therefore is necessarily
approximate and introduces free parameters. For a consistent treatment,
multidimensional hydrodynamical simulations are inevitable.

\section{Modeling deflagrations}

The internal structure of deflagrations in the dense core of
carbon-oxygen white dwarfs cannot be resolved in current
multidimensional explosion simulations. Therefore, modeling
approaches\index{flame models for deflagrations} are required. Two
general strategies can be distinguished that have been used in such
simulations: The first broadens the flame artificially by tuning the
microphysical transport coefficients and the progress of the reaction
such that a structure emerges that can be resolved on the scale of the
computational grid.  The second approach treats the flame as a sharp
discontinuity with no attempt to model its internal structure. It has
to be emphasized that neither of the approaches captures the physical
processes inside the flame correctly.

The approach of artificially broadened flames was introduced to the
field of thermonuclear supernova simulations by \cite{khokhlov1995a}
\cite[see also][]{vladimirova2006a, calder2007a}. It is based on a
reaction progress variable $\phi$ that governs energy release and
species conversion. The evolution of $\phi$ is described by an
advection-diffusion-reaction (ADR)
equation\index{advection-diffusion-reaction (ADR) flame model}
\begin{equation}
\frac{\partial \phi}{\partial t} + \mbf{v} \cdot \nabla \phi = \kappa
\nabla^2 \phi + \frac{1}{\tau} R(\phi),
\end{equation}
and hence the flame model is often referred to as ADR approach.  In the
absence of advection ($\mbf{v} = 0$), $\phi$ is propagated by
diffusion and produced by reaction. The diffusion coefficient $\kappa$
and the time scale of the reaction $\tau$ do not reflect microphysical
processes but are model parameters chosen such that the global flame
properties are obtained in the desired way: a model flame is defined
that matches the laminar flame propagation speed of the physical flame
but is broad enough to be resolved with a few cells on the
computational grid. A more sophisticated version of this flame model
involves individual progress variables for the different burning
stages in the flame \citep[carbon consumption, oxygen consumption and
  burning of silicon group elements to nuclear statistical
  equilibrium, see][]{calder2007a}.

The level-set approach\index{level-set approach}, in contrast, models
the flame as a discontinuity. Energy is released, and species are
converted instantaneously at the position of the discontinuity. Such a
discontinuity can be treated numerically by associating it to the
zero-level set of a signed distance function $G(\mbf{x}, t)$ ,
$|\nabla G| = 1$, such that the flame is modeled as the moving
hypersurface $\Gamma(t)$:
\begin{equation}
\Gamma (t) := \{\mbf{x} | G(\mbf{x}, t) = 0 \}
\end{equation}
\citep{osher1988a}. The propagation of this hypersurface is given by
the evolution of the signed distance function
\begin{equation}
\frac{\partial G}{\partial t} = (\mbf{v}_\mathrm{u} \mbf{n} +
s_\mathrm{u}) |\nabla G|,
\end{equation}
where $\mbf{v}_\mathrm{u}$, $\mbf{n}$, and $s_\mathrm{u}$ denote the
fluid velocity in the unburnt material, the normal vector to the flame
front, and the flame speed with respect to the fuel, respectively. The
$G$-function is defined to be negative in the fuel and positive in the
ashes \citep[for details of the implementation of the level-set
  approach for thermonuclear flames in white dwarf material
  see][]{reinecke1999a}. From $G$, the location of the flame can be
reconstructed with subgrid cell resolution.

Clearly, both approaches do not consistently treat the burning
microphysics but are models to propagate a flame-like structure on
scales resolvable in a thermonuclear supernova simulation. Therefore,
the model parameters have to be calibrated to reproduce properties of
the physical flames (in particular the flame speed and the energy
release). These have to be known independently and are taken from
one-dimensional microscopic flame simulations
\citep[e.g.][]{timmes1992a}. Both the ADR and the level-set approaches
are widely used to model deflagrations in thermonuclear supernova
explosion simulations. Although the ADR model may seem physically
better motivated, it has disadvantages: because the model flame has an
artificial finite width, it suffers from curvature effects close to
the grid scale, and it modifies fluid flows in neighboring cells that
are physically very far away from the actual flame. This may alter the
response of ADR model flames to instabilities and
turbulence. Minimizing such effects is one of the reasons why ADR
approaches are usually combined with adaptive mesh refinement (AMR)
techniques. The level-set approach, in contrast, is in principle able
to localize the flame front with subgrid-scale resolution as a
discontinuity (in the hydrodynamical variables, at least as much as the
employed hydrodynamics solver is able to represent discontinuities),
and AMR is usually not employed here.

As discussed in Sect.~\ref{sect:def_inst}, flames do not propagate
laminar in white dwarfs. They are subject to instabilities and
interaction with turbulent motions, which significantly enhances the
burning efficiency.  Such effects have to be taken into account in any
flame model for simulating thermonuclear supernova explosions.  This
is not automatically guaranteed because of the finite spatial
resolutions these simulations reach. Consequently, the numerical flame
model will lack structure from the unresolved scales and be
artificially smooth. For the flamelet regime, surface area enhancement
and burning acceleration due to flame-turbulence interaction are
therefore not fully reproduced in simulations. To compensate for the
effect of missing flame substructure, the model flame front is
propagated on the grid scale with an effective \emph{turbulent flame
  speed} $s_\mathrm{t}$ instead of the laminar value $s_\mathrm{l}$,
that applies only to unresolved microscopic scales. This is
critical for the success of many thermonuclear supernova models.

An approach used in some of the published thermonuclear supernova
simulations \citep[e.g.][]{gamezo2003a, jordan2008a, townsley2007a,
  jordan2012b} is to scale the effective turbulent flame
speed to the velocity of buoyantly rising bubbles
\begin{equation}
s_\mathrm{t} \propto \sqrt{g L \frac{\rho_\mathrm{u} -
    \rho_\mathrm{b}}{\rho_\mathrm{u} + \rho_\mathrm{b}}},
\end{equation}
where $g$ is the gravitational acceleration and $L$ denotes the length
scale of the bubble, here associated to the scale of the computational
grid. \cite{khokhlov1995a} discusses a
self-similarity in structures of turbulent flame fronts subject to
buoyancy as a motivation for this approach.

The assumption of turbulent flame speeds being set by buoyancy effects
has been called into question by \citet{ciaraldi2009a}, who argue that
a turbulence-driven speed should be used instead \citep[but
  see][]{hicks2015a}. An elaborate technique to account for
flame-turbulence interaction is based on turbulent subgrid-scale
models\index{subgrid-scale turbulence model}. According to
\cite{damkoehler1940a} the turbulent burning speed $s_\mathrm{t}$ of a
flame on a certain scale should be proportional to the turbulent
velocity fluctuations $v'$ on that scale. These are determined from a
subgrid-scale turbulence model in a Large-Eddy Simulation (LES)
approach. In an LES, only the largest scales of the turbulent cascade,
where energy is injected into it, are resolved together with the onset
of the inertial range. At the grid scale, the turbulent energy cascade
is cut off, and additional modeling is required at and below this
scale. Such turbulent subgrid-scale models were introduced by
\cite{niemeyer1995b}, \cite{schmidt2006b, schmidt2006c} to the field
of thermonuclear supernova explosion simulations \citep[see][for an
  alternative approach]{jackson2014a}. They are based on a balance
equation that describes the budget of turbulent kinetic energy on the
unresolved scales, with terms accounting for transport of kinetic
energy from the resolved to the unresolved scales, for energy
transport on the unresolved scales, for dissipation by viscosity on
unresolved scales, and for the action of the Archimedian force on
unresolved scales. These subgrid-scale models for turbulence-flame
interactions were used in several multidimensional simulations of
deflagrations in white dwarfs leading to thermonuclear supernova
explosions, e.g., \cite{reinecke2002b, reinecke2002d},
\cite{roepke2004c, roepke2005b}, \cite{roepke2005c},
\cite{roepke2006a, roepke2007c, roepke2007a}, and \cite{fink2014a}.

\section{Detonations}
\label{sect:det}

The physical mechanism of detonations\index{detonations} differs
fundamentally from that of deflagrations. A detonation propagates due
to the progression of a shock wave that compresses and heats fuel
material such that it starts to burn. This burning, in turn, releases
the energy to further support the shock wave. The temperature increase
enhances the reaction rate -- a process, however, that proceeds more
slowly because the shock structure is too thin to allow significant
reactions while passing. Behind this shock wave, reactions take place
and release energy leading to a further increase in temperature which
also raises the reaction rate. Fuel is consumed until depleted. Unlike
deflagrations, the propagation of detonations is not determined by
microphysical transport processes, but instead by hydrodynamical
effects. Consequently, these are fast combustion fronts advancing with
the speed of the shock wave supersonically with respect to the fuel
state.

\begin{figure}[t]
\centerline{\includegraphics[width=0.6\textwidth]{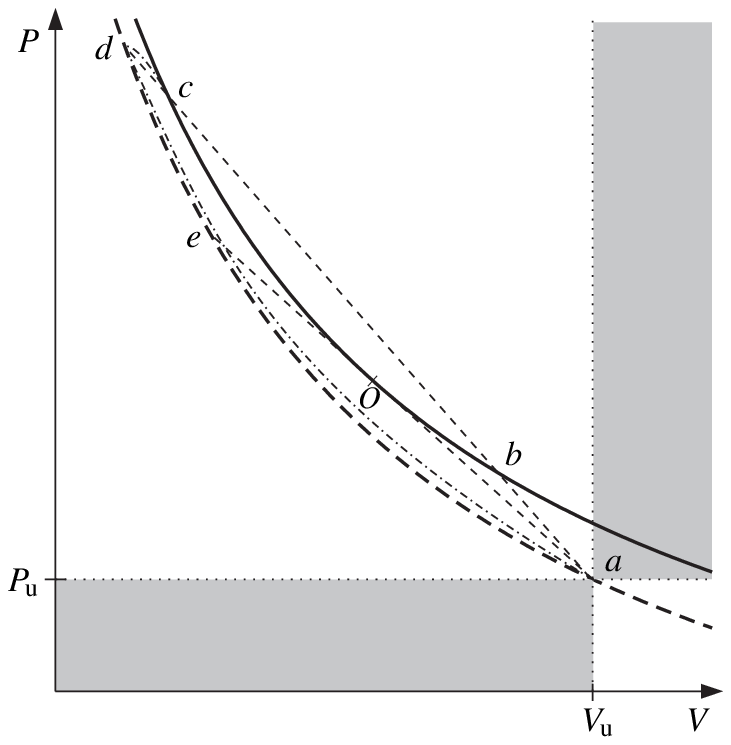}}
\caption{Close-up of the detonation branch. [See also
  \cite{landau1987a} and \cite{linan1993a}. Reprinted with minor
  modifications from \cite{roepke_phd}]\label{fig:det} }
\end{figure}

For planar detonations, a theoretical picture was independently
developed by \citet{zeldovich1940a}, \citet{vonneumann1942a}, and
\citet{doering1943a}. This so-called ZND model\index{detonations, ZND
  model} can be illustrated in the $(P,V)$-plane of
Fig.~\ref{fig:hugoniot}. Consider the detonation branch only (see
Fig.~\ref{fig:det}). The shock in the ZND model is assumed to be
infinitely thin as in the discontinuity approximation discussed in
Sect.~\ref{sect:fronts}. It evolves the initial state from point
\textit{a} along the detonation adiabat (bold dashed curve) over point
\textit{e} to state \textit{d}, which marks the highest pressure in
the detonation structure and is called the \emph{von Neumann spike.}
Here, reactions set in and lead to a heating and expansion of the
material. Contrary to the above-discussed discontinuity approximation,
the reactions are no longer assumed to be instantaneous but proceed
with finite rates behind the shock. The corresponding path toward the
final burnt state, e.g. \textit{c}, must follow down a Rayleigh line,
i.e.\ \textit{dc}, because mass and momentum are conserved for all
intermediate states. This path can be pictured as consisting of a
sequence of intermediate states in which adiabats corresponding to
partial but progressively more complete burning intersect with the
Rayleigh line until the detonation adiabat of the full energy release
is reached. Relaxing the assumptions of the ZND theory leads to a more
realistic path that is indicated as a dashed-dotted line in
Fig.~\ref{fig:det}.

A special case is obtained when the final state is located at the
Chapman-Jouguet point \textit{O}. The detonation then propagates
sonically with respect to the burnt material. Detonations often
propagate in this mode and attain a steady structure \citep[see][for a
  more comprehensive treatment of detonation
  theory]{fickett1979a}\index{detonations, Chapman-Jouguet}. In white
dwarf material of sufficiently high density to burn to nuclear
statistical equilibrium\index{nuclear statistical equilibrium},
detonations diverge from the Chapman-Jouguet type because of the
endothermic nature of the involved photodissociation reactions
\citep{khokhlov1989a, sharpe1999a, gamezo1999a}. They are of
pathological type\index{detonations, pathological} and travel with
greater speeds. At lower densities, however, the Chapman-Jouguet case
is a valid approximation.

\section{Multidimensional structure of detonations}

The detonation model outlined above describes the simplified
one-dimensional case of planar detonation waves. Because detonations
propagate into the fuel with supersonic speeds, they are not subject
to external hydrodynamical instabilities, such as the Rayleigh-Taylor
or the Kelvin-Helmholtz instabilities. Nonetheless, real detonations
possess a three-dimensional structure\index{detonations,
  multidimensional structure}. Planar detonations are unstable and
form incident shocks, transverse shock waves, and triple points. Their
continuous interactions cause the emergence of cellular patterns
\citep[e.g.][]{fickett1979a} that are observed in terrestrial
experiments\index{detonations, cellular instability}.

Transverse waves move back and forth along the detonation
perpendicular to its direction of motion.  At the points where they
collide, so-called triple points emerge. Between these triple points,
the shocks show significant curvature and are too weak to sustain the
detonation. In the triple points, the compression is stronger, and this
drives the reactions and the propagation of the detonation. The tracks
of these triple points form a cellular pattern in the downstream
material. This effect is observed in small-scale simulations of
thermonuclear detonations in white dwarf matter \citep{boisseau1996a,
  gamezo1999a, timmes2000e}. It alters the characteristics of the
detonation. Burning is more complete in the triple points, and thus the
chemical composition of the ashes is inhomogeneous behind the front
with pockets of less completely burned material off the paths of the
triple points. The emerging complex multidimensional detonation
structure is widened compared to the prediction of the one-dimensional
model, and its propagation speed is lower.

The one-dimensional theory therefore applies only in an average sense
and provides an acceptable approximation if multidimensional structure
cannot be resolved. According to \cite{gamezo1999a}, this is the case
for detonations in dense white dwarf material. At lower densities,
these structures may grow to sizes comparable to the scales of the
exploding star. Even if unresolved, the changes in propagation
velocity and ash composition due to the multidimensional structure may
affect the characteristics of thermonuclear supernova explosions,
although not by much \citep{timmes2000e}. Overall, multidimensional
detonations are weaker and may quench at higher densities than
expected from one-dimensional theory.

\section{Modeling detonations}

In many models of thermonuclear supernova explosions, detonations play
an important role. They seem to be required to produce the stratified
chemical composition observed in the outer layers of normal Type Ia
supernovae. Because the associated shock wave compresses the material,
they lead to more complete burning than deflagrations for the same
fuel density. To produce the intermediate mass elements observed in
the spectra of Type Ia supernovae, substantial burning has to take
place in low-density material. This is not given for detonations in
Chandrasekhar-mass white dwarfs in hydrostatic equilibrium. The high
densities in these would lead to an exclusive production of iron group
elements. Therefore, detonations have to trigger after a phase of
expansion caused by an initial deflagration in a Chandrasekhar-mass
white dwarf, or they lead to explosions of less compact
sub-Chandrasekhar mass configurations.

Contrary to the case of deflagrations, detonations do not rely on
microphysical transport, but arise directly from reactive fluid
dynamics. Therefore, no additional modeling of microphysical processes
is required for capturing these processes in numerical simulations
that are based on the reactive Euler equations.  In multidimensional
simulations of thermonuclear explosions of white dwarfs, however, it
is impossible to resolve the inner structure of detonations. Similar
to the case of deflagrations, two approaches have been taken: one that
broadens the structure to fit on the numerical grid and one that
treats detonations as sharp discontinuities\index{flame models for
  detonations}.

Broadened detonations arise naturally in a system modeled with the
reactive Euler equations, provided they are triggered and the
thermodynamic conditions are sufficient to allow for their
propagation. The correct jumps in the hydrodynamical quantities as
well as the detonation speed are retained for Chapman-Jouguet
detonations even if their microphysical structure is not resolved.
This means that contrary to broadened deflagrations, the detonation
speed is determined consistently in the model and does not have to be
provided externally. It is, however, necessary to artificially
suppress unphysical burning inside the (too wide) shock
structure. This approach has been employed in a number of
thermonuclear supernova simulations \citep[e.g.]{gamezo2005a,
  meakin2009a, townsley2009a, jackson2010a, jordan2012a, moll2013a,
  pakmor2013a}.
 
The level-set approach allows to numerically treat detonations as
discontinuities and has been employed in simulations of thermonuclear
supernovae \citep[e.g.][]{golombek2005a, roepke2007b, kasen2009a,
  sim2010a,fink2010a, pakmor2010a, pakmor2012a, seitenzahl2013a,
  ohlmann2014a, marquardt2015a, seitenzahl2016a}. Here, the material
crossed by the model detonation is instantaneously converted into
nuclear ash, and the corresponding energy is released. This does not
require to include an extensive reaction network as the details of the
burning are not resolved. Therefore, energy release, species
conversion, and detonation speed are parameters of the model that have
to be determined externally. Such a parametrization can lead to
inaccuracies, and, therefore, contrary to the case of deflagrations, it
may be preferred to use the broadened detonations approach. The
advantages of the level-set technique, however, are that unphysical
propagation of detonations over deflagration ash regions
\citep{maier2006a} in the delayed detonation model can easily be
prevented and that the speed of non-Chapman-Jouguet detonations can be
correctly set.

\section{Deflagration-to-detonation transitions}

The model of delayed detonations for Type Ia supernova explosions
\citep{khokhlov1991a} assumes a spontaneous transition of the burning
mode from an initial subsonic deflagration to a supersonic
detonation. Such transitions are indeed observed in terrestrial
combustion, where they are usually arise from the interaction of the
deflagration flame with walls or obstacles in the combustion
region. In the case of thermonuclear supernovae,
deflagration-to-detonation transitions
(DDTs)\index{deflagration-to-detonation transitions} have to take
place in an unconfined medium. It is unclear, however, whether such
unconfined DDTs occur in nature.

One possibility \citep[but see][for an alternative]{poludnenko2011a}
is a suitable spatial gradient of autoignition delay (also called
\emph{induction}) times of the reactions. Such a configuration may
lead to a coherent runaway of the reactions with a phase velocity that
is sufficient to ramp up to a detonation wave. The original idea of
\citet{zeldovich1970a}\index{Zeldovich gradient mechanism} was later
extended to the so-called shock wave amplification through
coherent energy release (SWACER) picture by \citet{lee1978a}\index{SWACER
  mechanism}.

This detonation initiation mechanism requires a preconditioning of the
fuel material with a shallow temperature gradient to arrange for the
required induction time gradient. It has been speculated that
sufficiently strong turbulence in a late phase of deflagration burning
in a thermonuclear supernova explosion can provide such conditions, in
particular once the fuel density has dropped to
$\lesssim$$10^7\,\mathrm{g}\,\mathrm{cm}^{-3}$ \citep{lisewski2000b,
  woosley2009a}, but it is difficult to identify such regions in
simulations of thermonuclear supernova explosions \citep[see][for a
  possibility]{roepke2007d}. Several simulations therefore
artificially prescribe the DDT spot or trigger the transition once the
deflagration flame reaches a certain density
threshold. \citet{ciaraldi2013a} propose a subgrid-scale model for
DDTs that takes into account turbulence properties. This model was
employed in the delayed detonation simulations of
\citet{seitenzahl2013a}.

\section{Conclusions}

The theory of combustion provides the basis for models of
thermonuclear supernova explosions. Because of its technological
application, combustion is well studied, and theory of the basic
phenomena has reached a rather mature state. It can therefore be
claimed that the physical principles of combustion wave propagation --
at least for the one-dimensional case -- are understood to the level
of precision needed for modeling the astrophysical
events. Complications arise in multidimensional models because of
instabilities and interaction with turbulence.

The numerical implementation of combustion waves remains challenging
because of the limited spatial resolution in multidimensional
supernova explosion simulations. Both deflagration and detonation
propagation modes can either be represented as an artificially
broadened structure fitting on the numerical grid or as a sharp
discontinuity between fuel and ash. Advantages and disadvantages of
both approaches have been discussed. It seems that for deflagration
modeling a discontinuity representation is favorable, while detonations
are better captured in the broadened wave approach, but this may
depend on the particular situation and explosion scenario under
consideration. Different possibilities exist for modeling the
interaction of deflagrations with instabilities and turbulence, and
elaborate subgrid-scale techniques have been employed in supernova
simulations. The multidimensional structure of detonations is not
recovered (nor accounted for) in current thermonuclear supernova
explosion simulations. Although it seems unlikely to have a
significant impact on the results of such simulations in terms of
ejecta composition and predicted observables, future model improvement
may require to account for this effect.

Deflagration-to-detonation transitions in unconfined media are an
unsolved problem in combustion theory. In delayed detonation models of
Type Ia supernovae, this adds an uncertainty. Although necessary
conditions for such a transition have been derived, it remains unclear
whether the mechanism is actually realized in nature.

Current combustion wave models are successful in providing a
qualitative picture of Type Ia supernova explosion scenarios. Future
efforts, however, are required to make them precise enough to allow
for a detailed quantitative comparison with observed supernova events.

\begin{acknowledgement}
This research was supported by the Munich Institute for Astro- and
Particle Physics (MIAPP) of the DFG cluster of excellence ``Origin and
Structure of the Universe.'' In particular, participation in the MIAPP
workshop ``The physics of supernovae,'' where this article was
initiated, is gratefully acknowledged.  The work of FKR is supported
by the Klaus-Tschira Foundation.
\end{acknowledgement}

\end{document}